\documentclass[11pt,letterpaper]{amsart}

\usepackage{amsmath,amsthm,amssymb}
\usepackage{xspace}

\renewcommand{\marginpar}[1]{\relax}

\newcommand{\fcn}[1]{({#1})}

\newcommand{\ignore}[1]{\relax}

\newcommand{\onlyconf}[1]{\relax}

\newtheorem{theorem}{Theorem}
\newtheorem{lemma}[theorem]{Lemma}

\newtheorem{example}[theorem]{Example}
\newtheorem{definition}[theorem]{Definition}

\def\R{{\mathbb R}}

\newcommand{\pcsp}{PCSP\xspace}
\newcommand{\pcsps}{PCSPs\xspace}
\newcommand{\pcsplong}{Polynomial CSP\xspace}
\newcommand{\pcspslong}{Polynomial CSPs\xspace}
\newcommand{\rcsp}{RCSP\xspace}
\newcommand{\rcsps}{RCSPs\xspace}
\newcommand{\rcsplong}{RCSP\xspace}

\newcommand{\tcsp}{2-\csp}
\newcommand{\csp}{CSP\xspace}
\newcommand{\csps}{CSPs\xspace}

\newcommand{\gbsx}[1]{\relax}
\newcommand{\disjunion}{\uplus}

\newcommand{\kset}{[k]}

\newcommand{\Ostar}[1]{{O^\star\!}\left(#1\right)}
\newcommand{\invimage}{\sigma^{-1}}

\newcommand{\stilde}{\tilde{s}}
\newcommand{\s}{l}
\newcommand{\Itilde}{\widetilde{I}}
\newcommand{\allv}{(\forall v \in V) \qquad}
\newcommand{\alle}{(\forall xy \in E) \qquad}
\newcommand{\comma}{\; , \;}

\newcommand{\Istar}{I^*}
\newcommand{\ptilde}{{\widetilde p}}
\newcommand{\Gtilde}{{\widetilde G}}
\newcommand{\sett}[1]{\{ {#1} \}}

\renewcommand{\a}{\alpha}
\renewcommand{\b}{\beta}
\renewcommand{\c}{\gamma}
\newcommand{\D}{\Delta}
\newcommand{\tr}{\operatorname{tr}}
\newcommand{\parens}[1]{\Big( {#1} \Big)}
\newcommand{\weight}{\operatorname{weight}}
\newcommand{\q}{\ptilde}
\newcommand{\Gbar}{\overline{G}}

\begin{document}

\bibliographystyle{amsalpha}

\title[\pcsplong]
{
Polynomial Constraint Satisfaction Problems, Graph \\ 
Bisection, and the Ising Partition Function
}

\author[Alexander D. Scott]{Alexander D. Scott$^*$}
\thanks{$^*$ Research supported in part by EPSRC grant GR/S26323/01}%
\address[Alexander D. Scott]{
Mathematical Institute\\
University of Oxford\\
24-29 St Giles'\\ 
Oxford, OX1 3LB, UK}
\email{scott@maths.ox.ac.uk}

\author{Gregory B. Sorkin%
}
\address[Gregory B. Sorkin]{
Department of Mathematical Sciences \\
IBM T.J.\ Watson Research Center \\
Yorktown Heights NY 10598, USA}
\email{sorkin@watson.ibm.com}

\begin{abstract}
We introduce a problem class we call Polynomial Constraint Satisfaction
Problems, or PCSP.  
Where the usual CSPs from computer science and optimization
have real-valued score functions, 
and partition functions from physics have monomials, 
PCSP has scores that are arbitrary multivariate formal polynomials,  
or indeed take values in an arbitrary ring.

Although PCSP is much more general than CSP, remarkably, 
all (exact, exponential-time) algorithms we know of for 2-CSP 
(where each score depends on at most 2 variables) 
extend 
to 2-PCSP, 
at the expense of just a polynomial factor in running time 
for adding and multiplying polynomials rather than reals.  
Specifically, four algorithms that extend are: 
the reduction-based algorithm of Scott and Sorkin; 
the specialization of that approach to sparse random instances, 
where the algorithm runs in polynomial expected time; 
dynamic-programming algorithms based on tree decompositions; 
and the split-and-list matrix-multiplication algorithm of Williams.

This gives the first 
polynomial-space exact algorithm 
more efficient than exhaustive enumeration
for the well-studied problems of finding a minimum bisection of a graph, 
and calculating the partition function of an Ising model. 
Furthermore, PCSP solves both optimization and counting versions 
of a wide range of problems, including all CSPs, 
and thus enables samplers including uniform sampling 
of optimal solutions and Gibbs sampling of all solutions. 
\end{abstract}

\maketitle

\section{Introduction}

We introduce a problem class we call Polynomial 2-CSP,
which generalizes more common 2-CSP constraint satisfaction problems,
yet is solvable by 
adaptations of all the best CSP algorithms 
we know of, with essentially equal efficiency.
This provides the first efficient polynomial-space 
exact algorithm for the 
well-studied problem of finding the maximum bisection of a graph
(exponential-space algorithms were previously known),
and the first exact algorithm of any sort,
other than naive exhaustive enumeration,
for the equally important problem of
computing the partition function of an Ising model.
It also solves a breadth of other optimization problems beyond
the class 2-CSP, including some quite convoluted ones,
in a simple, uniform framework.
It allows counting and sampling as well as optimization,
and again, does this in time bounds matching those for 2-CSP.
Let us elaborate.

Many decision and optimization problems, 
such as 3-colorability, 
Max Cut, Max $k$-Cut, and Max 2-Sat, belong to the class Max \tcsp of 
constraint satisfaction problems with at most two variables per constraint.  
A Max \tcsp (or for short, simply \csp) instance 
defines a ``score function'' or ``soft constraint'' on each 
vertex and edge of a ``constraint graph'',
and its solution is a vertex coloring or ``assignment'' 
maximizing the total score.

We define a more general class we call Polynomial 2-CSP,
or simply 2-\pcsp or \pcsp.
(The \pcsp definition extends obviously to longer clauses,
but much less is known about algorithms even for 3-CSP.) 
Comparing with a CSP,
the vertex and edge scores of a \pcsp are 
(potentially multivariate) polynomials 
(rather than real numbers),
the score of a coloring is the product of 
all vertex and edge scores
(rather than their sum),
and an instance's ``value'', or ``partition function'', 
is the sum 
of scores over all colorings
(rather than the maximum).
{In fact the scores may be ``generalized polynomials'', 
whose powers are not restricted to positive integers but may be
arbitrary reals,
but we use ``polynomial'' for want of a better term.
}

Compared with the ``max--sum'' CSP formulation typical in the CSP community,
PCSP has a ``sum--product'' form 
commonly seen as the ``partition function'' in statistical physics:
a sum over configurations, of a product of exponentials of ``energies''.
In the physics settings we are familiar with, 
the partition function is always a product of monomials
(sometimes multivariate, for example with energy and magnetization),
whereas our \pcsp{s} permit polynomials. 
With general polynomials,
\pcsp is closed under a variety of operations 
enabling its (relatively) efficient solution by 
a number of algorithmic approaches.

\pcsp includes many optimization problems not in the class \csp,
including Maximum Bisection, Maximum Clique,
Sparsest Cut, various judicious partitioning problems,
Max Ones 2-Sat, and many others.
Capitalizing on the fact that a \pcsp's score functions
may be {multivariate} polynomials,
in all these cases,
we simply track several score functions at once.
For example, for Max Bisection we track
the size of one partition as well as the size of the cut.

The $H$-coloring of a graph $G$ has received much recent attention,
including the book \cite{Hell}, 
the application to soft constraints in \cite{Bulatov},
and the application to asymmetric hard constraints in \cite{Dyer}.
$H$-coloring can be encoded as a \pcsp by giving every edge 
of $G$ the same (symmetric) score function.  
However, \pcsp can also give a different (and asymmetric) score function
to each edge, and indeed this is an essential feature 
as it leaves the class closed under various reductions.

\pcsps give a powerful technique for 
{counting and sampling} solutions. 
Given a \csp, 
there is a natural way to obtain a corresponding \pcsp.
Solving the \pcsp gives a generating function for the number of \csp
assignments of each possible score, 
which allows us to count or (by successively solving subinstances) 
to sample uniformly from solutions of maximum score, 
or to sample all solutions according to the Gibbs distribution 
or any other score-based distribution.  

What makes \pcsp so interesting is that, 
despite being more general than \csp,
all the best \csp algorithms we know of
can 
be extended to solve \pcsp,
with the same running-time bounds.
(Exact \csp algorithms are generally exponential-time,
and solving a \pcsp takes just a polynomial factor longer than 
solving a \csp instance of the same size,
for the extra overhead of adding and multiplying polynomials
rather than reals.)
Specifically, four algorithms that extend are: 
the split-and-list matrix-multiplication algorithm of Williams \cite{Williams};
the reduction-based algorithm of Scott and Sorkin \cite{faster};
the specialization of that approach to sparse random instances,
where the algorithm runs in polynomial expected time \cite{linear}; 
and
dynamic-programming algorithms based on 
tree decompositions \cite{Karpinski,faster,Kneis4}.

Both the definition of \pcsp and all our algorithms only require 
us to take sums and products of polynomial ``scores''.  
It is thus natural to work in a still more general class that we call 
Ring \csp, 
where each score function and the ``value'' of an instance 
are given by element of an arbitrary ring~$R$; 
\pcsps are a special case where $R$ is a polynomial ring.

\section{Outline}
Following a small amount of notation in Section~\ref{notation},
Section~\ref{pcspdef} defines the classes \csp, \pcsp and \rcsp. 
Section~\ref{examples} provides examples of \csp{s} and
related \pcsps; because this includes many problems beyond \csp,
such as 
graph bisection, the Ising partition function, and 
judicious partitioning,
it is a focal point of the paper.  

In Sections~\ref{sec:Williams}--\ref{treewidth},
we show how four \csp algorithms extend to \pcsp/\rcsp,
and analyze them in terms of the number of
ring operations (or sums and products of polynomials in the case of \pcsps) that they require. 
In Section~\ref{sec:ring} we consider the time and space
complexity of the ring operations for \pcsps. 
We put the two together in Section~\ref{conclusions},
summarizing the time and space complexity of each algorithm,
and noting the circumstances favoring one algorithm over another.

In addition to counting maximum \csp solutions, 
a \pcsp's partition function also enables construction of
an optimal solution, 
and various forms of perfect random sampling from 
optimal solutions or all solutions;
this is taken up in 
Section~\ref{sampling}.

\section{Notation} \label{notation}
In the next section we will define the class of \csp{s}
and our new class \pcsp of \pcspslong. 
An instance of either has a ``constraint graph'' $G=(V,E)$
with vertex set $V$ and edge set~$E$,
and we reserve the symbols $G$, $V$ and $E$ for these roles.
An instance of \csp or \pcsp also has a domain of 
\emph{values} or \emph{colors} that may be 
assigned to the vertices (variables), for example
$\{\text{true},\text{false}\}$ for a satisfiability problem, 
or a set of colors for a graph coloring problem. 
In general we will denote this domain by~$\kset$, 
interpreted as $\{0,\dots,k-1\}$ or (it makes no difference)
$\{1,\dots,k\}$.
At the heart of both the \csp and \pcsp instance 
will be cost or ``score'' terms $s_v\fcn i$ ($v \in V(G)$, $i \in \kset$)
and $s_{xy}\fcn {i,j}$ ($x,y \in V(G)$, $i,j \in \kset$). 

For \csp{s} the scores $s$ are real numbers,
while for \pcsps they are polynomials which we
write in the variable $z$ if univariate, 
or if multivariate over 
$z,w$ or $z,w_1,w_2,\dots$.
Just as \csp scores $s$ need not be positive integers, 
exponents in the \pcsp score ``polynomials'' may be
fractional or negative (or both).
(The term ``fractional polynomial'' is used in the literature of
statistical regression.)

We use $\disjunion$ to indicate disjoint union, 
so $V_0 \disjunion V_1 = V$ means that $V_0$ and $V_1$ partition~$V$,
i.e., $V_0 \cap V_1 = \emptyset$ and $V_0 \cup V_1 = V$.
The notation $O^\star$ hides polynomial factors in any parameters,
so for example $O(n 2^{5+19m/100})$ is contained in $\Ostar{2^{19m/100}}$.

\section{\csp, \pcsplong, and Ring \csp} \label{pcspdef}

\subsection{CSP}
Let us begin by defining the problem class \csp over a domain of
size~$k$.  An \emph{instance $I$ of \csp with constraint graph $G=(V,E)$
and domain $\kset$} has the following ingredients:
\begin{enumerate}
\item a real number $s_\emptyset$;
\item for each vertex $v\in V$ and color $i\in\kset$,
a real number~$s_v\fcn i$;
\item for each edge $xy\in E$ and any colors $i,j \in \kset$,
a real number~$s_{xy}\fcn {i,j}$.  
\end{enumerate}
We shall refer to these quantities as, respectively, the \emph{nullary
score}, the \emph{vertex scores}, and the \emph{edge scores}.  
Note that we want only one score 
for an edge $\sett{i,j}$ with given colors $\sett{x,y}$ 
assigned its respective endpoints, so
$s_{xy}\fcn {i,j}$ and 
$s_{yx}\fcn {j,i}$ are taken to be equivalent names for the 
same score (or one may simply assume that~$x<y$).

Given an \emph{assignment} (or \emph{coloring}) 
$\sigma: V\to \kset$, 
we define the \emph{score of $\sigma$} to be the real number
$$I(\sigma):=s_\emptyset+\sum_{v\in V}s_v\fcn {\sigma(v)}
 +\sum_{xy\in E}s_{xy}\fcn {\sigma(x),\sigma(y)}.$$
The instance's solution is the maximum possible score
\begin{align}
\max_{\sigma: V \to \kset} I(\sigma) 
 &=
\max_{\sigma: V \to \kset} 
 \Big(
 s_\emptyset+\sum_{v\in V}s_v\fcn {\sigma(v)}
 +\sum_{xy\in E}s_{xy}\fcn {\sigma(x),\sigma(y)} 
 \Big) 
 \label{maxsum}
\end{align}
or the assignment $\sigma$ achieving the maximum; 
it is straightforward to get either from the other.
For obvious reasons, \eqref{maxsum} is commonly referred to
as a \emph{max--sum} formulation. 

In this paper we take a generating-function approach,
among other things enabling us 
to count the number of solutions satisfying various properties.
Given an instance~$I$ of \csp, the corresponding generating function is the
polynomial
\begin{equation}\label{gf}
\sum_{\sigma:V\to\kset}z^{I(\sigma)}.
\end{equation}
More generally, if we want to keep track of several quantities
simultaneously, say $I(\sigma), J(\sigma),\cdots$, we can consider a
multivariate generating function
$\sum_{\sigma}z^{I(\sigma)}w^{J(\sigma)}\cdots$.  Calculating the
generating function in the obvious way, by running through all
$k^{|V|}$ assignments, is clearly very slow, so it is desirable to
have more efficient algorithms.

In working with generating functions, we borrow some notions from
statistical physics.  We think of the score $I(\sigma)$ as a
``Hamiltonian'' measuring the ``energy'' of a configuration $\sigma$.
Thus edge scores correspond to ``pair interactions'' between
adjacent sites, while vertex scores measure the effect of a
``magnetic field''. (From this perspective, the nullary score is just a
constant that disappears after normalization.)  The generating
function is then the ``partition function'' for this model.
(The partition function is often written in the form 
$\sum_\sigma \exp(-\beta I(\sigma))$, where $\beta$ is 
known as the \emph{inverse temperature},  
but substituting $z$ for $e^{-\beta}$ yields the partition
function in polynomial form.)

A crucial element of our approach 
is that the score $I(\sigma)$ can be broken up as a
sum of local interactions, and thus an expression such as $z^{I(\sigma)}$
can be expressed as a product of monomials corresponding to local
interactions:
$$z^{I(\sigma)}=z^{s_\emptyset}\cdot\prod_{v\in
V}z^{s_v(\sigma(v))}\cdot\prod_{xy\in
E}z^{s_{xy}\fcn {\sigma(x),\sigma(y)}}.
$$
In order to provide a framework for this approach, we
introduce a generalized version of constraint satisfaction where the
scores are polynomials (in some set of variables) instead of real
numbers, and the score of an assignment is taken as a product rather
than a sum.

\subsection{Polynomial \csp}

An \emph{instance $I$ of \pcsplong, with constraint graph $G$
and domain~$\kset$}, has the following ingredients:
\begin{enumerate}
\item a set of formal variables, 
and various polynomials over these variables:
\item a polynomial $p_\emptyset$;
\item for each vertex $v\in V$ and color $i\in \kset$,
a polynomial $p_v\fcn i$;
\item for each edge 
$xy \in E$ and pair of colors $i,j \in \kset$, 
a polynomial~$p_{xy}\fcn {i,j}$.
\end{enumerate}
We refer to these three types of polynomial as, respectively,
the \emph{nullary polynomial}, the \emph{vertex polynomials}, and the
\emph{edge polynomials}.  
(Reiterating from Section~\ref{notation}, 
we use ``polynomial'' in a general sense, 
allowing negative and fractional powers.)
We want only one polynomial for a given edge with given colors on its 
endpoints, so again we either take 
$p_{xy}\fcn {i,j}$ and $p_{yx}\fcn {j,i}$ to be equivalent 
or simply assume that $x<y$.

Given an assignment $\sigma: V\to \kset$, we define the
\emph{score of $\sigma$} to be the polynomial
\begin{align}
I(\sigma) & :=
 p_\emptyset\cdot
 \prod_{v\in V}p_v\fcn {\sigma(v)}\cdot
  \prod_{xy\in E}p_{xy}\fcn {\sigma(x),\sigma(y)} .
  \label{pscore}
\end{align}
We then define the \emph{partition
function $Z_I$ of $I$} by
\begin{align}
Z_I
 &=
 \sum_{\sigma \colon V\to\kset} I(\sigma)
 =
 \sum_{\sigma \colon V\to\kset}
  p_\emptyset\cdot
  \prod_{v\in V}p_v\fcn {\sigma(v)}\cdot
  \prod_{xy\in E}p_{xy}\fcn {\sigma(x),\sigma(y)} .
  \label{sumprod}
\end{align}
The partition function is the ``solution'' of a \pcsp instance.
In contrast with the max--sum formulation of \eqref{maxsum},
\eqref{sumprod}~is referred to as a \emph{sum--product} formulation.
The sense of the PCSP formulation will 
become clearer with examples in the next section.

\subsection{Ring \csp}

The definition of $I(\sigma)$ in equation \eqref{pscore}, 
and for the most part our analysis in later sections,
requires only addition and multiplication of scores.  
It therefore makes sense to work in a still more general context 
where the score functions are elements of an arbitrary ring.  
We define an 
\emph{instance $I$ of \rcsplong over a ring $R$, with constraint graph $G$
and domain~$\kset$}, to have the following ingredients:
\begin{enumerate}
\item a ring element $r_\emptyset$;
\item for each vertex $v\in V$ and color $i\in \kset$,
a ring element $r_v\fcn i$;
\item for each edge 
$xy \in E$ and pair of colors $i,j \in \kset$, 
a ring element~$r_{xy}\fcn {i,j}$.
\end{enumerate}
The \emph{score} of an assignment $\sigma: V\to \kset$ 
is the ring element
$$I(\sigma):=r_\emptyset\cdot\prod_{v\in
V}r_v\fcn {\sigma(v)}\cdot\prod_{xy\in
E}r_{xy}\fcn {\sigma(x),\sigma(y)},$$ 
and the \emph{partition
function $Z_I$ of $I$} is
$$Z_I=\sum_{\sigma \colon V\to\kset}I(\sigma).$$
Thus \pcsps are the special case of \rcsps where the ring $R$ is a polynomial ring over the reals.

\section{Examples}\label{examples}

In this section we show how some standard
problems can be written as \pcsps.
First we show that for every \csp 
there is a naturally corresponding \pcsp;
then we illustrate how various problems, including some that are not \csps,
can be expressed as \pcsps.
We will turn to algorithms in later sections, 
but for now bear in mind that a \pcsp over a constraint graph $G$
can be solved about as efficiently as a \csp over the same graph.

\subsection{Generating function of a simple CSP}\label{gfscsp}

\begin{definition}[Generating function of an instance] \label{genfn}
Given an instance $I$ of \csp on a constraint graph $G=(V,E)$ and
variable domain~$\kset$, we can define a corresponding
instance $\Istar$ of \pcsp with the same graph
and variable domain, and polynomials
\begin{align*}
 & p_\emptyset =z^{s_\emptyset}
\\
(\forall v \in V ,\: \forall i \in \kset) \qquad &
p_v\fcn i =z^{s_v\fcn i} 
\\
(\forall xy \in E ,\:  \forall i,j \in \kset) \qquad &
p_{xy}\fcn {i,j} =z^{s_{xy}\fcn {i,j}} .
\end{align*}
\end{definition}

The connection between $I$ and $\Istar$ is given by the following simple
observation.

\begin{lemma} \label{genfnlemma}
Let $I$ be an instance of \csp, and let $\Istar$ be the corresponding
\pcsp instance.  
Then the partition function $Z_{\Istar}$ is the
generating function \eqref{gf} for the instance~$I$.
\end{lemma}

\begin{proof}
For any assignment $\sigma$, we have
\begin{align*}
\Istar(\sigma)
&=z^{s_\emptyset}
\cdot\prod_{v\in V}z^{s_v\fcn i}
\cdot\prod_{xy\in E}z^{s_{xy}\fcn {i,j}}
\\ &
=z^{s_\emptyset+\sum_{v\in V}s_v\fcn i+\sum_{xy\in E}s_{xy}\fcn{i,j}}
\\ &
=z^{I(\sigma)}.
\end{align*}
It therefore follows that the partition function 
$$Z_{\Istar}
 =\sum_\sigma {\Istar(\sigma)}
 =\sum_\sigma z^{I(\sigma)}$$
is the generating function for the original constraint satisfaction problem.
\end{proof}
Similar results are easily seen to hold 
for generating functions in more than one variable.

\subsection{Max Cut and Max Dicut}

Max Cut provides a simple illustration 
of Definition~\ref{genfn} and Lemma~\ref{genfnlemma}.  
Let us first write Max Cut as a \csp; 
we will then construct the corresponding \pcsp.

\begin{example}[Max Cut CSP]
Given a graph~$G=(V,E)$, set $k=2$ and
define a \csp instance $I$ by 
\begin{align*}
 & s_\emptyset=0\\
 (\forall v \in V) \qquad & s_v\fcn 0=s_v\fcn 1=0 \\
 (\forall xy\in E) \qquad & 
s_{xy}\fcn {0,1}=s_{xy}\fcn {1,0} = 1 \comma
s_{xy}\fcn {0,0}=s_{xy}\fcn {1,1}=0 .
\end{align*}
\end{example}
With $\sigma^{-1}(i)=\{v \colon \sigma(v)=i\}$, 
note that $(V_0,V_1) = (\invimage(0),\invimage(1))$ is a partition of~$V$, 
and 
\begin{align}
I(\sigma) 
 &= \sum_{\substack{xy \in E \colon \\ \sigma(x)=0,\: \sigma(y)=1}} 1 
 = e(V_0,V_1) 
\label{enotation}
\end{align}
is the size of the cut induced by~$\sigma$.
The corresponding \pcsp instance is obtained as in Definition~\ref{genfn}.

\begin{example}[Max Cut \pcsp]
Given a graph~$G=(V,E)$, we set $k=2$ and
define a \pcsp instance $I$ by 
\begin{align*}
 & p_\emptyset=1 \\ 
\allv & p_v\fcn 0=p_v\fcn 1 =1 \\
\alle & p_{xy}\fcn {0,1}=p_{xy}\fcn {1,0}=z \quad , \quad
p_{xy}\fcn {0,0}=p_{xy}\fcn {1,1}=1 .
\end{align*}
\end{example}
\noindent
(In all such cases, a $1$ on the right hand side may be thought of as $z^0$.)

By Lemma~\ref{genfnlemma},
the partition function $Z_I$ is therefore the generating function for cuts:
\begin{equation}\label{mc0}
Z_I=\sum 2c_iz^i,
\end{equation}
where $c_i$ is the number of cuts of size~$i$,
and the factor $2$ appears because each cut $(V_0,V_1)$ 
also appears as $(V_1,V_0)$.
The size of a maximum cut is the degree of~$Z_I$, and the number of
maximum cuts is half the leading coefficient.  

Note that the partition function \eqref{mc0} 
is the partition function of the Ising model with no external field 
(see below for a definition).  
Thus we have recovered the familiar fact that, 
up to a change of variables, 
the partition function of the Ising model is the generating function for cuts.

We can also encode weighted instances of Max Cut with
edge weights 
$w:E\to\R$ by modifying the third line of the
definition above to $$p_{xy}\fcn {0,1}=p_{xy}\fcn {1,0}=z^{w(xy)} \qquad \forall
xy\in E.$$

Similarly, we can encode weighted Max Dicut (maximum directed cut) by
setting
\begin{align*}
 & p_\emptyset =1\\
\allv & p_v\fcn 0=p_v\fcn 1 =1 \\
\alle & 
p_{xy}\fcn {0,1} =z^{w(xy)} \comma
p_{xy}\fcn {1,0} =z^{w(yx)} \comma
p_{xy}\fcn {0,0}=p_{xy}\fcn {1,1} =1 ,
\end{align*}
where $w(xy)$ denotes the weight of the \emph{directed} edge~$xy$, 
and we define $w(xy)=0$ if there is no such edge.
Max $k$-Cut is encoded the same way,
only with $k$-valued variables in place of binary ones.

\subsection{The Ising model and Max Bisection}
A slight generalization of the previous example 
allows us to handle the Ising model.

\begin{example}[Ising CSP model]\label{ising}
The Ising model with edge weights $J$ and external field $h$ on a
graph $G$ is defined in terms of its Hamiltonian~$H$. For an
assignment $\sigma:V\to\{0,1\}$, we define
$$H(\sigma)=
J\sum_{xy\in E}\delta(\sigma(x),\sigma(y))
+h\sum_{v\in V}\sigma(v).$$
\end{example}

Here $\delta(a,b)$ is the delta function, returning
0 if $a=b$, and 1 otherwise,
$J$ is the \emph{interaction strength},
and $h$ is the \emph{external magnetic field}.
In analogy with \eqref{enotation}, taking
$V_i = \invimage(i)$,
we may rewrite $H$ as
\begin{align*}
H(\sigma) = J e(V_0,V_1) + h |V_1| .
\end{align*}
Note that $H$ is an instance of (plain) \csp. 

The \emph{partition function} of the Ising model at 
\emph{inverse temperature}~$\beta$ is
\begin{align}
Z_{\rm Ising}
&=\sum_\sigma e^{-\beta H(\sigma)}
 = \sum_{V_0 \disjunion V_1=V} w^{|V_1|} z^{e(V_0,V_1)},
 \label{IsingSub}
\end{align}
where we have written $w=e^{-\beta h}$ and $z=e^{-\beta J}$,
and the last sum is taken over 
ordered pairs $(V_0,V_1)$ that partition~$V$. 
With this change of variables, the Ising partition function is 
easily expressed as partition function of a \pcsp, 
this time over two variables.

\begin{example}[Ising \pcsp] Define an Ising \pcsp instance $I$ by
\begin{align*}
 & p_\emptyset =1\\
\allv & p_v\fcn 0 =1 \\
\alle & p_v\fcn 1 =w \comma
p_{xy}\fcn {0,1}=p_{xy}\fcn {1,0} =z \comma
p_{xy}\fcn {0,0}=p_{xy}\fcn {1,1} =1 .
\end{align*}
\end{example}

With our usual notation, 
$I(\sigma)
=w^{|V_1|} z^{e(V_0,V_1)}$, and so 
the partition function for this \pcsp is
equal to~\eqref{IsingSub}.

More generally, the Potts model can be written in a similar way.  
Less generally, the zero magnetization ($h=0$) anti-ferromagnetic ($J=-1$)
special case of this Ising model has 
maximum cuts as its ``ground states'' (lowest-energy states): 
taking $\beta \to \infty$ means that these states 
dominate the sum comprising~$Z_{\rm Ising}$.
To view it another way, 
the maximum degree in $z$ of $Z(w,z)$ 
is the size of a \emph{maximum cut} in~$G$. 

The Ising \pcsp can also be used to handle \emph{Max Bisection}.
This is important, because most \csp algorithms 
(other than dynamic programming)
can solve Max Cut, but cannot be applied to Max Bisection because 
there is no way to force them to generate a balanced cut. 
The Ising \pcsp does this by tracking two variables at once: 
the bisections of $G$ correspond to terms in $Z_{\rm Ising}$ 
that have degree $\lfloor n/2\rfloor$ in~$w$.  
(Each bisection is counted once if $n$ is odd, twice if $n$ is even.)
Extracting these terms gives the generating function for bisections of~$G$.

The same partition function also yields a \emph{sparsest cut}
of the graph: sparsest cuts correspond to terms 
with the largest ratio of the 
power of $z$ (number of cut edges) to the power of $w$ 
(number of vertices in one partition).

\subsection{Max Independent Set and Max Clique}
Maximum Independent Set (MIS) is easily expressed as a \csp: 
\begin{example}[MIS CSP]
\begin{align*}
 & s_\emptyset=0 \\
\allv & s_v\fcn 0 =0 \comma s_v\fcn 1 =1 \\
\alle &
s_{xy}\fcn {0,0}=s_{xy}\fcn {0,1}=s_{xy}\fcn {1,0}=0  \comma s_{xy}\fcn {1,1}={-2} .
\end{align*}
\end{example}
Maximum clique%
\footnote{We use clique in the sense of a complete subgraph,
not the stronger definition as a \emph{maximal} complete subgraph.%
}
cannot be modeled in the same way, 
because the clique constraint is enforced by non-edges, 
which are not an element of the model.
Of course a maximum clique in $G$ corresponds to a 
maximum independent set in its complement graph $\Gbar$, 
but for our purposes this can be very different, 
as we typically parametrize running time in terms of the number of edges,
and a Max Clique instance on a sparse graph $G$ with $|E|$ edges
becomes an MIS instance on the dense graph $\Gbar$
with $\binom n2 - |E|$ edges.
We discuss this further in Section~\ref{keyresults}.
However, as with Max Bisection, 
it is possible to model a sparse Max Clique instance as a \pcsp 
(with the input graph $G$ as its constraint graph)
by introducing a second variable.
\begin{example}[Max Clique \pcsp] \label{clique}
\begin{align*}
 & s_\emptyset =1 \\
\allv & s_v\fcn 0 =1 \comma s_v\fcn 1 =w \\
\alle & 
s_{xy}\fcn {0,0}=s_{xy}\fcn {0,1}=s_{xy}\fcn {1,0}=1  \comma s_{xy}\fcn {1,1} =z .
\end{align*}
\end{example}

An assignment $\sigma$ has score $I(\sigma) = w^{|V_1|} z^{e(G|_{V_1})}$, 
the power of $w$ counting the number of vertices in the chosen set 
$V_1=\sigma^{-1}(1)$, 
and the power of $z$ counting the number of edges induced by that set.
In the partition function, 
$k$-cliques correspond to terms of the form $w^k z^{\binom k 2}$. 
Of course, independent sets of cardinality $k$ correspond 
to terms with $w^k z^0$: 
the \pcsp simultaneously counts maximum cliques and maximum independent sets 
(among other things).

The same approach also counts maximum-weight cliques and/or independent sets, 
in a graph with vertex and/or edge weights,
if we introduce a third formal variable $w'$ and set the scores as before
except for
$s_v(1) = w {w'}^{\weight(v)}$ and
$s_{xy}(1,1) = z {w'}^{\weight(x,y)}$.
Cliques correspond to terms $w^k z^{\binom k 2} {w'}^{\weight}$,
and the maximum-weight clique the term with the largest power of~$w'$.

\subsection{Judicious partitions}
Similar techniques may be applied to various \emph{judicious partitioning}
problems \cite{BScomb,Sbcc}, 
such as finding a cut of a graph which minimizes
$\max\{e(V_0),e(V_1)\}$.

\begin{example} [Judicious bipartition \pcsp] 
\begin{align*}
 & s_\emptyset =1\\
 \allv & s_v\fcn 0=s_v\fcn 1 =1 \\
 \alle & s_{xy}\fcn {0,1}=s_{xy}\fcn {1,0} =1 \comma
  s_{xy}\fcn {0,0} =z_1 \comma
  s_{xy}\fcn {1,1} =z_2 .
\end{align*}
\end{example}

A simple calculation shows that $I(\sigma)=z_0^{e(V_0)}z_1^{e(V_1)}$, 
where $V_i=\sigma^{-1}(i)$.  
Thus the problem of minimizing $\max\{e(V_0),e(V_1)\}$ 
(or, for instance, the problem of maximizing $\min\{e(V_0),e(V_1)\}$) 
can be solved by examining the partition function.  

We might further ask for a \emph{bisection} 
that minimizes $\max\{e(V_0),e(V_1)\}$, 
and this is easily achieved by introducing a third variable: 
we set $s_v\fcn 0=w$ and examine only terms of the partition function 
that have degree $\lfloor |V|/2\rfloor$ in $w$.

\subsection{Simultaneous assignments}
We can think of the foregoing example, 
of finding a balanced bisection minimizing $\max\{e(V_0),e(V_1)\}$,
as a case of having more than one CSP on a single set of variables.
In other examples, 
we might have two instances of Max Sat that we wish 
to treat as a bi-criterion optimization problem;
or a Max Sat instance to optimize subject to satisfaction of 
a Sat instance; 
or we might wish to find a partition of a vertex set yielding 
a large cut for two different graphs on the same vertices.
All such problems are now straightforward: use a variable $z$ to
encode one problem and a variable $w$ to encode the second.  
(Thus, the initial edge scores are of the form 
$z^{\rm first score}w^{\rm second score}$.)

\section{Split-and-list, matrix-multiplication algorithm}
\label{sec:Williams}
The algorithmic approach introduced by Williams \cite{Williams},
which he calls ``split and list'',
is characterized by its use of fast matrix multiplication.
For binary 2-CSPs it runs in time $\Ostar{2^{\omega n/3}}$,
where $\omega$ is the ``matrix multiplication exponent''
such that two $n \times n$ matrices can be multiplied together 
in time $O(n^\omega)$.
It also requires exponential space, $O(2^{2n/3})$.
The best known bound on the matrix multiplication constant
is $\omega < 2.37\ldots$, 
from the celebrated algorithm of
Coppersmith and Winograd \cite{Coppersmith-Winograd},
which works over arbitrary rings.

Koivisto \cite{Koivisto} extended Williams' algorithm from 
CSP (with a max--sum formulation)
to counting-CSP (with sum--product formulation),
and Koivisto's version transfers immediately to the PCSP and RCSP context.
While the sum--product formulation is common in statistical physics, 
Koivisto's use of it appears to be one of the few besides the present work 
in the field of exact algorithms.
For Williams' algorithm, the sum--product formulation is not only more general
but also simpler.

\subsection{Algorithm}
Koivisto's counting
extension of Williams' algorithm is very much in the spirit
of the original,
and itself works in the PCSP and RCSP settings.
Arbitrarily partition the variables $V$ into 3 equal-sized sets, 
$A$, $B$ and $C$.
Explicitly list every possible assignment $\a: A \mapsto [k]$
of variables in~$A$ 
and similarly every assignment $\beta$ for $B$ and $\gamma$ for~$C$.
Explicitly construct $M_{AA}$, 
a $k^{|A|} \times k^{|A|}$ diagonal matrix with elements
\begin{align*}
M_{AA}(\a,\a) & :=
 \prod_{v\in A} p_v\fcn {\a(v)} \cdot
 \prod_{xy\in E \cap \sett{A \times A}} p_{xy}\fcn {\a(x),\a(y)} 
 ,
\end{align*}
and likewise $M_{BB}$ and $M_{CC}$.
Also explicitly construct $M_{AB}$,
a $k^{|A|} \times k^{|B|}$ matrix with elements
\begin{align*}
M_{AB}(\a,\b) :=
 \prod_{xy\in E \cap \sett{A \times B}} p_{xy}\fcn {\a(x),\b(y)} 
 ,
\end{align*}
and likewise $M_{BC}$ and $M_{CA}$.
Finally, explicitly calculate
\begin{align*}
M &= p_\emptyset M_{AA} M_{AB} M_{BB} M_{BC} M_{CC} M_{CA} ,
\end{align*}
the multiplication by $p_\emptyset$ being a scalar product
and the others matrix products.
Triples $(\a,\b,\c)$ are in one to one correspondence
with assignments $\sigma: V \mapsto [k]$, 
so the terms of the trace of $M$,
\begin{align*}
\tr(M) &=
\sum_\a \sum_\b \sum_\c p_\emptyset 
 M_{AA}(\a,\a) M_{AB}(\a,\b)
 M_{BB}(\b,\b) M_{BC}(\b,\c)
 M_{CC}(\c,\c) M_{CA}(\c,\a) ,
\end{align*}
are in one to one correspondence with the terms of $I(\sigma)$,
merely distinguishing 
whether each monadic score is associated with $A$, $B$, or $C$,
and each dyadic score with 
$A\times A$, $A \times B$, etc.
Thus, the partition function is the trace,
\begin{align*}
Z_I &= \tr(M) .
\end{align*}

\subsection{Complexity}
Given that two $n \times n$ matrices over an arbitrary ring
can be multiplied using $O(n^\omega)$ ring operations,
we have established the following result.
\typeout{Fast matrix mult must be linear space,
or else the space bound might equal the time bound.}

\begin{theorem}
Let $R$ be a ring and let $G$ be a graph with $n$ vertices.  
Let $I$ be any \rcsp over $R$, with constraint graph $G$ and domain $\kset$.  
Then the algorithm above calculates the partition function $Z_I$ 
with $O\left(k^{\omega n/3}\right)$ ring operations.
\end{theorem}

For \pcsps, as we discuss in Section~\ref{sec:ring},
under modest conditions each ring operation takes time $\Ostar1$,
that is, time polynomial in the input size.  
In this case, the algorithm will
run in time and space $\Ostar{k^{\omega n/3}}$.
(Space $O(k^{2 n/3})$ suffices if the fast matrix multiplication
algorithm uses linear space.)

\subsection{A remark}
This version of the algorithm is simpler than Williams' because
matrix multiplication provides exactly the sum-of-products
that RCSP calls for,
not the maximum-of-sums needed for the usual 2-CSP.
To get the latter,
Williams combines the monadic scores with the dyadic ones
(this would correspond to incorporating $M_{AA}$ into $M_{AB}$);
``guesses'' an optimal score $s$ and how it is partitioned as
$s = s_{AB}+s_{BC}+s_{CA}$ 
among $A \times B$, $B \times C$, and $C \times A$;
defines zero-one matrices so that
$M_{AB}(\a,\b)=1$ if the assignments $\a$ and $\b$
produce the desired value $s_{AB}$, and 0 otherwise;
and multiplies these binary matrices to find (and count) triangles, 
which correspond to assignments $(\a,\b,\c)$ yielding $s_{AB}+s_{BC}+s_{CA}$.
Williams' original algorithm must iterate over all guesses,
of which there are $\Theta(m^3)$ even for 0--1 score functions,
while no such guessing is involved in the sum--product version.

\section{A reductive algorithm}\label{reductions}
The preponderance of algorithms for CSPs such as Max 2-Sat
work by reduction:
they reduce an input instance to one or more ``smaller'' instances
in such a way that the solutions to the latter, found recursively,
yield a solution to the former.
The fastest polynomial-space algorithm for general Max 2-CSP
is the reductive algorithm of \cite{faster},
running in time $\Ostar{k^{19m/100}}$ for an instance with $m$
2-variable constraints and domain~$\kset$.
While we will not describe that paper's
``Algorithm~B'' in detail,
if an instance $I$ has (a constraint graph with)
a vertex of degree 0, 1, or 2, 
a Type 0, 1, or 2 reduction (respectively) replaces $I$ 
with an equivalent, smaller instance~$I'$.
Otherwise, a Type~3 reduction replaces
$I$ with $k$ smaller instances
$I_0,\ldots,I_{k-1}$, 
the largest of whose solutions is the solution to~$I$.
If a Type~3 reduction results in a disconnected graph, 
the components are solved separately and summed.

The algorithm extends to \pcsp (and indeed \rcsp), 
the key point being that the Type 0, 1, 2, and~3 \csp reductions 
have \pcsp (and \rcsp) analogues.
By working with polynomials rather than real numbers
we are able to carry substantially more information: 
for instance, 
the old \csp reductions correspond to 
the new \pcsp reductions with the polynomials truncated 
to their leading terms.

\subsection{Complexity}
We exhibit the extended reductions below,
where it will also be evident that each can be performed using 
$O(k^3)$ ring operations.  
The extension of \cite{faster}'s Algorithm~B to RCSPs
can then be seen to satisfy the following.

\begin{theorem} \label{reductive}
Let $R$ be a ring and let $G$ be a graph with $n$ vertices and $m$ edges.  
Let $I$ be any \rcsp over $R$, with constraint graph $G$ and domain $\kset$.  
Then the extended Algorithm~B calculates the partition function $Z_I$ 
in polynomial space and
with $\Ostar{k^{19m/100}}$ ring operations.  
\end{theorem}

\subsection{The reductions}
A Type~0 reduction expresses the partition function of an \rcsp 
as a product of partition functions of two smaller instances 
(or in an important special case, a single such instance).
Type 1 and 2 reductions each equate the generating function of 
an instance to that of an instance with one vertex less.
Finally, a Type 3 reduction produces $k$ instances, whose
partition functions sum to the partition function of the original instance.  
In each case, once the reduction is written down, verifying
its validity is just a matter of checking a straightforward identity.

A word of intuition, deriving from the earlier \csp reductions, 
may be helpful (though some readers may prefer to go straight to 
the equations).
A \csp 2-reduction was performed on a vertex $v$ of degree~2, 
with neighbors $u$ and $w$. 
The key observation was that in a maximization problem, 
the optimal assignment $\sigma(v)$ is a function of the 
assignments $\sigma(u)$ and $\sigma(w)$. 
That is,
given any assignments $\sigma(u)=i$ and $\sigma(w)=j$, 
the optimal total $\stilde_{uw}\fcn {i,j}$ of the scores
of the vertex $v$, the edges $uv$ and $vw$,
and the edge $uw$ (if present, and otherwise taken to be the 0 score function)
is 
\begin{align}
\stilde_{uw}\fcn {i,j} &= 
 \max_\s \big\{ 
   s_{uw}\fcn {i,j} + s_{uv}\fcn {i, \s} + s_v(\s) + s_{vw}\fcn {\s, j}  
 \big\}
 \label{csp} .
\end{align}
By deleting $v$ from the \csp instance and replacing the original 
score $s_{uw}$ with the score~$\stilde_{uw}$,
we obtain a smaller instance with the same maximum value.
A 2-reduction
in the \rcsp context 
works the same way,
except that the max-of-sums of \eqref{csp} 
is replaced by the sum-of-products of \eqref{pcsp2}.

\medskip

\noindent{\bf Type~0 Reduction.}
Suppose $I$ is an \rcsp instance whose constraint graph $G$ is disconnected.
Let $V=V_1\cup V_2$ be a nontrivial partition such
that $e(V_1,V_2)=0$.  Let $I_1$ and $I_2$ be the subinstances obtained
by restriction to $V_1$ and $V_2$, except that we define the nullary
scores by
\begin{align*}
p_\emptyset^{(I_1)}&=p_\emptyset^{(I)}
\comma
p_\emptyset^{(I_2)}=1.
\end{align*}
A straightforward calculation shows that, 
for any assignment $\sigma:V\to\kset$, we have
\begin{align*}
I(\sigma)
&=p_\emptyset\cdot\prod_{v\in V}p_v\fcn {\sigma(v)}\cdot\prod_{xy\in E}p_{xy}\fcn {\sigma(x),\sigma(y)}\\
&=\left(p_\emptyset\cdot\prod_{v\in V_1}p_v\fcn {\sigma(v)}\cdot\prod_{xy\in E(G[V_1])}p_{xy}\fcn {\sigma(x),\sigma(y)}\right)
\left(1\cdot\prod_{v\in V_2}p_v\fcn {\sigma(v)}\cdot\prod_{xy\in E(G[V_2])}p_{xy}\fcn {\sigma(x),\sigma(y)}\right)\\
&=I_1(\sigma_1)I_2(\sigma_2),
\end{align*}
where $\sigma_i$ denotes the restriction of $\sigma$ to $V_i$.  
It follows easily that
$$Z_I=Z_{I_1}Z_{I_2}.$$
Thus in order to calculate $Z_I$ it suffices to calculate $Z_{I_1}$
and $Z_{I_2}$.  

When one component of $G$ is an isolated vertex $v$, with $V'=V\setminus v$,
one term becomes trivial:
$Z_I=\big(p_\emptyset\cdot\sum_{i\in\kset}p_v\fcn i\big)\cdot Z_{I_2}.$
So in this case 
$$Z_I=Z_{\Itilde},$$
where $\Itilde$ is
the instance obtained from $I$ by deleting~$v$, defining
$$p_\emptyset\fcn {\Itilde}=p_\emptyset\cdot\sum_{i\in\kset}p_v\fcn i$$
(a trivial calculation),
and leaving all other scores unchanged.
This reduces the computation of the partition function of $I$ 
to that of an instance one vertex smaller.

\medskip

\noindent{\bf Type~1 Reduction.}
Suppose that $I$ is an instance with constraint graph~$G$, 
and $v\in V$ has degree~$1$.  
Let $w$ be the neighbor of~$v$.
We shall replace $I$ by an ``equivalent'' instance $\Itilde$ 
(one with the same partition function)
with constraint graph $G\setminus v$.

We define the instance $\Itilde$ 
by giving $w$ vertex scores
\begin{equation}\label{t1def}
{\ptilde}_w\fcn i=p_w\fcn i\sum_{j=0}^{k-1}(p_{wv}\fcn {i,j}\cdot p_v\fcn j).
\end{equation}
All other scores remain unchanged 
except for $p_v\fcn *$ and $p_{vw}\fcn {*,*}$, 
which are deleted along with~$v$.

To show that $Z_I=Z_{\Itilde}$, let $\sigma:V\setminus v\to\kset$
be any assignment and, for $j\in \kset$, extend $\sigma$ to 
$\sigma^j:V\to\kset$ defined by
$\sigma^j(v)=j$ and  $\sigma^j|_{V\setminus v}=\sigma$.  
Using~\eqref{t1def}, we have
\begin{align*}
\Itilde(\sigma)
&=\ptilde_\emptyset\cdot\prod_{x\in V\setminus v}\ptilde_x\fcn {\sigma(x)}\cdot\prod_{xy\in E(G\setminus v)}\ptilde_{xy}\fcn {\sigma(x),\sigma(y)}\\
&=\left(p_\emptyset\cdot\prod_{x\in V\setminus \{v,w\}} p_x\fcn {\sigma(x)}\cdot\prod_{xy\in E(G\setminus v)} p_{xy}\fcn {\sigma(x),\sigma(y)}\right)
\cdot \left(p_w\fcn {\sigma(w)}\cdot\sum_{j=0}^{k-1}p_{wv}\fcn {\sigma(w),j}\cdot p_v\fcn j\right)\\
&=\sum_{j=0}^{k-1}p_\emptyset\cdot\prod_{v\in V}p_v\fcn {\sigma^j(v)}
\cdot\prod_{xy\in E}p_{xy}\fcn {\sigma^j(x),\sigma^j(y)} \\
&=\sum_{j=0}^{k-1}I(\sigma^j) 
\end{align*}
and so
\begin{align*}
Z_I &=
\sum_{\sigma:V\to\kset}I(\sigma) \\ &=
\sum_{\sigma:V\setminus v\to\kset}\sum_{j=0}^{k-1}I(\sigma^j) \\ &=
\sum_{\sigma:V\setminus v\to\kset}\Itilde(\sigma) \\ &=
Z_{\Itilde} .
\end{align*}

\medskip

\noindent{\bf Type~2 Reduction.}
Suppose that $I$ is an instance with constraint graph~$G$, and $v\in
V$ has degree~$2$.  
Let $u$ and $w$ be the neighbors of $v$ in~$G$.  
We define an instance $\Itilde$ with constraint graph~$\Gtilde$, 
which will have fewer vertices and edges than~$G$.
$\Gtilde$ is obtained from
$G$ by deleting $v$ and adding an edge $uw$ (if the edge is not
already present).  
By analogy with the plain CSP 2-reduction from \cite{faster}
and recapitulated above around \eqref{csp},
we define $\Itilde$ by setting, for 
$i,j \in \kset$,
\begin{align}
\ptilde_{uw}\fcn {i,j}
 &= p_{uw}\fcn {i,j}\cdot\sum_{\s=0}^{k-1}p_{uv}\fcn {i,\s}p_{v}\fcn {\s}p_{vw}\fcn {\s, j},
 \label{pcsp2}
\end{align}
where we take $p_{uw}\fcn {i,j}=1$ if 
edge $uw$ was not previously present.
All other scores remain unchanged
except for $p_v\fcn *$, $p_{vu}\fcn {*,*}$ and $p_{vw}\fcn {*,*}$, 
which are deleted.

To show that $Z_I=Z_{\Itilde}$, 
let $\sigma:V\setminus v\to\kset$ be any assignment.  
As before, we write
$\sigma^\s$ for the assignment with 
$\sigma^\s|_{V\setminus v}\equiv\sigma$ and $\sigma^\s(v)=\s$.  
Then
\begin{align*}
\Itilde(\sigma)
&=
\ptilde_\emptyset
 \cdot\prod_{x\in V\setminus v}\ptilde_x\fcn {\sigma(x)}
 \cdot\prod_{xy\in E(\Gtilde)}\ptilde_{xy}\fcn {\sigma(x),\sigma(y)}
\\ &= 
p_\emptyset \prod_{x\in V\setminus v} p_x\fcn {\sigma(x)}
 \cdot\prod_{xy\in E(G\setminus v)} p_{xy}\fcn {\sigma(x),\sigma(y)}
 \cdot\sum_{\s=0}^{k-1} 
   p_{uv}\fcn {\sigma(u),\s}p_{v}\fcn {\s}p_{vw}\fcn {\s,\sigma(w)}
\\ &=
\sum_{\s=0}^{k-1} p_\emptyset 
 \prod_{x\in V\setminus v} p_x\fcn {\sigma^\s(x)}
 \cdot\prod_{xy\in E(G\setminus v)} p_{xy}\fcn {\sigma^\s(x),\sigma^\s(y)}
\\ &=
\sum_{\s=0}^{k-1}I(\sigma^\s).
\end{align*}
As with Type~1 reductions, this implies that $Z_I=Z_{\Itilde}$.

\medskip

\noindent{\bf Type~3 Reduction.}
Suppose that $I$ is an instance with constraint graph~$G$, 
and $v\in V$ has degree $3$ \emph{or more}.
Let $U$ be the set of neighbors of $v$ in~$G$.  
We define $k$ instances $\Itilde_0,\ldots,\Itilde_{k-1}$ 
each with constraint graph 
$\Gtilde=G\setminus v$.  
For $i \in \kset$, the $i$th instance $\Itilde_i$ 
corresponds to the set of assignments where we take $\sigma(v)=i$.

We define nullary scores for $\Itilde_i$ by setting
$$\q_\emptyset^{(i)}=p_\emptyset\cdot p_v\fcn i$$
and, for each neighbor $u \in U$ of $v$, and each $j \in \kset$,
vertex scores
$$(\q^{(i)})_{u}\fcn j = p_{u}\fcn j\cdot p_{ {u} v}\fcn {j,i}.$$
All other scores remain unchanged from $I$ 
except for $p_v\fcn *$ and $p_{v*}\fcn {*,*}$, 
which are deleted along with~$v$.

For any assignment $\sigma:V\setminus v\to\kset$, 
and $i \in \kset$
we write (as usual) $\sigma^i:V\to\kset$ 
for the assignment with $\sigma^i|_{V\setminus v}\equiv\sigma$ 
and $\sigma^i(v)=i$.  
Then, writing $W=V\setminus(v\cup\Gamma(v))$,
\begin{align*}
\Itilde_i(\sigma)
 &=
\ptilde_\emptyset
 \cdot \prod_{x\in V \setminus v} \ptilde_x\fcn {\sigma(x)}
 \cdot \prod_{xy\in E(G \setminus v)} \ptilde_{xy}\fcn {\sigma^i(x),\sigma^i(y)}
\\ &=
\parens{p_\emptyset p_v\fcn {i}}
 \cdot \parens{ 
    \prod_{x\in W}p_x\fcn {\sigma(x)}
    \cdot \prod_{u \in U} 
      \big[ p_u\fcn {\sigma(u)} p_{uv}\fcn {\sigma(u),i} \big]
 }
 \cdot \parens{ \prod_{xy\in E(G\setminus v)} p_{xy}\fcn {\sigma(x),\sigma(y)} }
\\ &=
p_\emptyset\prod_{x\in V}p_x\fcn {\sigma^i(x)}
\cdot\prod_{xy\in E}p_{xy}\fcn {\sigma^i(x),\sigma^i(y)}
\\ &=
I(\sigma^i) .
\end{align*}
Then
\begin{align*}
Z_I  
&=
\sum_{\sigma:V\to\kset}I(\sigma) 
\\ &=
\sum_{\sigma:V\setminus v\to\kset}\sum_{i=0}^{k-1}I(\sigma^i) 
\\ &=
\sum_{\sigma:V\setminus v\to\kset}\sum_{i=0}^{k-1}\Itilde_i(\sigma) 
\\ &=
\sum_{i=0}^{k-1}\sum_{\sigma:V\setminus v\to\kset}\Itilde_i(\sigma) 
\\ &=
\sum_{i=0}^{k-1}Z_{\Itilde_i}.
\end{align*}
Thus the partition function for $I$ 
is the sum of the partition functions for the~$\Itilde_i$.

\medskip

We can also write down another reduction, generalizing the Type~0
reduction above, although we will not employ it here.

\medskip

\noindent{\bf Cut Reduction.}
Suppose that $I$ is an instance with constraint graph~$G$, and let
$V_0\subset V$ be a vertex cut in~$G$.  Let $G_1,\ldots,G_r$ be the
components of $G\setminus V_0$.  We can calculate $Z_I$ by dividing
assignments into classes depending on their restriction to $V_0$.
Indeed, for each assignment $\sigma_0:V_0\to\kset$, let us define
$$I_0=p_\emptyset\cdot\prod_{v\in V_0}p_v\fcn {\sigma_0(v)}\prod_{xy\in E(G[V_0])}p_{xy}\fcn {\sigma_0(x),\sigma_0(y)},$$
and $r$ instances $I_1^{\sigma_0},\ldots,I_r^{\sigma_0}$ where $I_t^{\sigma_0}$ has score polynomials
\begin{align*}
 & \q_\emptyset =1\\
 \forall v\not\in V_0\qquad & \q_v\fcn i =p_v\fcn i\cdot\prod_{w\in V_0}p_{wv}\fcn {\sigma_0(w),i} \\
 \forall vw\in E[G\setminus V_0]\qquad & \q_{vw}\fcn {i,j} =p_{vw}\fcn {i,j} , \\
\end{align*}
where we take $p_{wv}\fcn {i,j}=1$ if $wv\not\in E$.

Then for any $\sigma:V\to\kset$, we have
$$I(\sigma)=I_0(\sigma|V_0)\cdot \prod_{i=1}^r I_i^{\sigma|V_0}(\sigma|V_i)$$
and so
$$
Z_I=\sum_{\sigma_0:V_0\to\kset}I_0(\sigma_0)\prod_{i=1}^r Z_{I_i^{\sigma_0}}.
$$
Thus the partition function $Z_I$ is the sum of $|V_0|^k$ products: if
$|V_0|$ is small, this provides an effective reduction to smaller
cases.  Note that this is what we have done in the Type~2 reduction
above, where we also use the fact that, for a cutset of size two, the
partition functions for the subinstance consisting of the single
vertex $v$ can be encoded using constraints on its neighbors.
Unfortunately, this is not in general possible for larger cutsets.

\section{An algorithm for sparse semi-random instances}
An algorithm closely related to the \csp Algorithm~B referenced above,
and using the same reductions, 
solves sparse boolean CSP instances in expected linear time \cite{linear}. 
Specifically, for $\lambda=\lambda(n)\ge0$ and a random graph $G(n,c/n)$ 
with $c \leq 1+ \lambda n^{-1/3}$
(below the giant-component threshold or in its so-called scaling window),
a boolean CSP instance $I$ with constraint graph $G$
is solved in expected time $O(n) \exp(1+\lambda^3)$.

\subsection{Complexity}
Because the reductions extend to \rcsp,
it follows the algorithm extends immediately to boolean \rcsp instances on such a constraint graph. 

\begin{theorem} \label{random}
Let $\lambda=\lambda(n)\ge0$.  For $c \leq 1+ \lambda n^{-1/3}$, 
let $G\in \mathcal{G}(n,c/n)$ be a random graph, 
and let $I$ be any boolean \rcsp over any ring $R$ with constraint graph $G$.  
Then the algorithm above calculates the partition function $Z_I$ with an expected $O(n) \exp(1+\lambda^3)$ ring operations.  
\end{theorem}

\section{Dynamic programming for graphs of small treewidth}
\label{treewidth}

Roughly, a \emph{tree decomposition} of a graph $G=(V,E)$ consists
of a tree $T$ each of whose vertices 
is associated with a ``bag'' $B \subseteq V$, 
such that every vertex of $G$ appears in some bag, 
every edge of $G$ has both endpoints in some bag,
and the set of bags containing any given vertex $v$ of $G$
forms a subtree of~$T$. 
The ``width'' of a tree decomposition is 
one less than the cardinality of the largest bag,
and the treewidth of $G$ is the minimum width of any tree decomposition.
For our purposes we will assume that a tree decomposition is given, 
though for graphs of constant-bounded treewidth,
a minimum-treewidth decomposition can be found 
in linear time \cite{robertsonpath,Reed,Bodlaender}.

Efficient algorithms for various sorts of constraint satisfaction
and related problems on graphs of small treewidth
have been studied since at least the mid-1980s,
with systematic approaches dating back at least 
to \cite{Dechter87,Dechter89,Arnborg}. 
A special issue of Discrete Applied Mathematics
was devoted to this and related topics in 1994 \cite{DAM},
and the field remains an extremely active area of research.

A recent series of results on treewidth is 
Monien and Preis's proof that every cubic graph with $m$ edges has 
bisection width at most $(1/6+o(1))m$ \cite{Monien},
Fomin and H{\o}ie's use of this to show that the pathwidth of
a cubic graph is bounded by $(1/6+o(1))m$ \cite{fomin},
and the use of this to prove that any graph 
has treewidth at most $(13/75+o(1))m$ \cite{faster,Kneis4}. 

Focusing on two problems of particular interest to us, 
as a key part of an approximation result 
in Jansen, Karpinski, Lingas and Seidel's \cite{Karpinski},
it was shown that given a graph $G$ 
and a tree decomposition $T$ of width~$b-1$,
an exact maximum (or minimum) bisection of $G$ can be found in time 
\emph{and space} $\Ostar{2^b}$,
using dynamic programming on the tree decomposition.
To find a maximum \emph{clique} in time $\Ostar{2^b}$ is even easier, 
and may be done in polynomial space,
using the well-known fact that any clique of $G$ is 
necessarily contained in a single bag of~$T$.

Here, we generalize these results for bisection and clique to 
the entire class \rcsp.
While the 
extension is reasonably straightforward,
the details are confusing enough that it is valuable to have the
general result available in this neatly packaged form,
and not have to apply the approach (or wonder if it can be applied)
to a particular problem.

\subsection{Complexity}
\begin{theorem} \label{thm:treewidth}
Let $G$ be a graph with $n$ vertices, $m$ edges,
and a tree decomposition $T$ of width~$b-1$.
Let $I$ be any \rcsp instance over $G$.
Given $I$ and $T$,
the partition function $Z_I$ can be calculated 
in space $O(k^{b-1})$ and
with $\Ostar{k^b}$ ring operations.
\end{theorem}

We sketch the proof before detailing it.
We express an assignment score 
as a product of ``bag scores'', $f_B$ for a bag~$B$.
In computing the partition function, we 
reduce on any leaf bag $B_1$ of $T$ and its parent~$B_2$, 
in a way closely related to the Type~1 reductions of Section~\ref{reductions}:
variables appearing \emph{only} in $B_1$ can be ``integrated out''
to yield a function $f'_{B_1}$ on a smaller set of variables.
By the definition of a tree decomposition,
that smaller set is a subset of the variables of~$B_2$,
so we can absorb $f'_{B_1}$ into $f_{B_2}$
by defining $f'_{B_2} = f_{B_2} \cdot f'_{B_1}$.
Deleting $B_1$ from $T$, and forgetting $f_{B_1}$,
completes the reduction.
We now present this more formally.

\begin{proof}
We first show how to express the score of 
an assignment as a product of ``bag scores''.
\newcommand{\bb}{\mathcal{B}}
\newcommand{\set}[1]{\left\{#1\right\}}
Let $\bb$ be the collection of all bags. 
While a vertex is typically found in several bags,
associate each vertex $v \in V(G)$ with just one bag that contains it;
similarly, associate each edge $e \in E(G)$ with 
some bag containing both its endpoints.
For a bag $B$, let $V(B)$ be the set of associated vertices 
and $E(B)$ the set of associated edges,
so that 
$\disjunion_{B \in \bb} V(B) = V(G)$ and
$\disjunion_{B \in \bb} E(B) = E(G)$.
\newcommand{\fargs}[1]{\sigma|_{#1}}
Define each bag score as the product of its associated 
vertex and edge scores:
\begin{align}
f_B(\fargs B) 
&= 
 \prod_{v \in V(B)} p_v\fcn {\sigma(v)} \cdot
 \prod_{xy \in E(B)} p_{xy}\fcn {\sigma(x), \sigma(y)} .
 \label{bagscore}
\end{align}
(Note that for an edge $xy \in E(B)$, the endpoints
$x$ and $y$ need not be in $V(B)$,
but they must be in $B$ itself,
and thus $f_B(\fargs B)$ is well defined,
depending only on assignments of vertices in~$B$.)
It is clear that for the original \pcsp instance~$I$,
\begin{align}
Z_I & :=
\sum_{\sigma \colon V \to \kset} 
 p_\emptyset\cdot
  \prod_{v\in V}p_v\fcn {\sigma(v)}\cdot
  \prod_{xy\in E}p_{xy}\fcn {\sigma(x),\sigma(y)}
\notag \\ &=
 \sum_{\sigma \colon V \to \kset} 
  p_\emptyset\cdot
  \prod_{B \in \bb} f_B(\fargs B). \label{bagform}
\end{align}

We now show a reduction that will preserve the form of~\eqref{bagform},
but with one bag fewer.
After reduction, the bag scores themselves will no longer depend simply
on vertex and edge scores as in~\eqref{bagscore},
which is why we work with \eqref{bagform} instead.

\newcommand{\vone}{V'}
\newcommand{\vrest}{V''}
\newcommand{\sone}{\sigma'}
\newcommand{\srest}{\sigma''}
Let $B_1$ be a leaf bag and $B_2$ its parent. 
Let $\vone = B_1 \setminus B_2$.
By the nature of a tree decomposition,
$\vone$ is the set of vertices found exclusively in $B_1$.
Let $\vrest = V \setminus \vone$.
Correspondingly, we will use
$\sone$ and $\srest$ for assignments from these sets to~$\kset$,
and $(\sone,\srest)$ for a complete assignment from $V$ to $\kset$.
In the following rewriting of~\eqref{bagform},
lines \eqref{deff1p} and \eqref{deff2p} are implicit definitions
of the functions $f'_{B_1}$ and~$f'_{B_2}$.
\begin{align} 
Z_I & =
 \sum_{\sigma \colon V \to \kset} 
  p_\emptyset \cdot
  \left( \prod_{B \in \bb \setminus \set{B_1,B_2}} f_B(\sigma''|_B) \right)
  \cdot
  f_{B_2}(\sigma|_{B_2}) \cdot
  f_{B_1}(\sigma|_{B_1}) 
  \label{prered}
\\ &=
 \sum_{\srest \colon \vrest \to \kset} 
  p_\emptyset \cdot
  \left( \prod_{B \in \bb \setminus \set{B_1,B_2}} f_B(\srest|_B) \right)
  \cdot
  f_{B_2}(\srest |_{B_2}) \cdot
 \sum_{\sone \colon \vone \to \kset} 
  f_{B_1}( (\srest,\sone) |_{B_1}) 
  \notag
\\ &=
 \sum_{\srest \colon \vrest \to \kset} 
  p_\emptyset \cdot
  \left( \prod_{B \in \bb \setminus \set{B_1,B_2}} f_B(\srest|_B) \right)
  \cdot
  f_{B_2}(\srest |_{B_2}) \cdot
  f'_{B_1}(\srest |_{B_1 \cap B_2}) 
  \label{deff1p}
\\ &=
 \sum_{\srest \colon \vrest \to \kset} 
  p_\emptyset \cdot
  \left( \prod_{B \in \bb \setminus \set{B_1,B_2}} f_B(\srest|_B) \right)
  \cdot
  f'_{B_2}(\srest |_{B_2}) .
  \label{deff2p}
\end{align}
This reduces the number of bags by one, as desired.
Note that the definition implicit in \eqref{deff1p},
that for any $\srest \colon \vrest \to \kset$,
$$ f'_{B_1}( (\srest) |_{B_1 \cap B_2}) 
 := 
 \sum_{\sone \colon \vone \to \kset} 
  f_{B_1}( (\srest,\sone) |_{B_1}) ,
$$
is well defined because
the right-hand side depends only on values of $(\srest,\sone)$
on the set $B_1 \cap B_2$.
Likewise, the definition implicit in \eqref{deff2p},
that for any $\srest \colon \vrest \to \kset$,
$$ 
  f'_{B_2}(\srest |_{B_2})
 := 
  f_{B_2}(\srest |_{B_2}) \cdot
  f'_{B_1}( (\srest) |_{B_1 \cap B_2}) ,
$$
is well defined because
the right-hand side depends only on values of $(\srest,\sone)$
on the set $B_2$.

In going from \eqref{prered} to \eqref{deff2p} we reduced the 
decomposition tree size by~1.
To do so we represented the functions $f'_{B_1}$ and $f'_{B_2}$ explicitly, 
writing down respectively $k^{|\vone|}$
and $k^{|V(B_2)|}$ score values.
The reduction does not change the size of any bag,
so if initially each bag has size at most $b-1$,
each reduction takes space $O(k^{b-1})$ 
and requires $\Ostar{k^{b-1}}$ ring operations
(each bag evaluation may require evaluating polynomially many
vertex and edge scores).

The algorithm really does require extra space $O(k^{b-1})$ 
to represent the updated score function $f'_{B_2}$,
because  $f'_{B_2}$ incorporates $f_{B_1}$ ``integrated out''
over $\sone \colon \vone \to \kset$
and can no longer be viewed as coming from a bag of vertex and edge scores.
We conclude that altogether the algorithm takes
space $\Ostar{k^{b-1}}$ and uses $\Ostar{k^{b-1}}$ ring operations.
\end{proof}

\section{Controlling the ring-operation complexity}
\label{poly}
\label{sec:ring}

In this section, we restrict to \pcspslong, 
and consider the complexity of performing the ring operations in our algorithms.
In general, \pcsp instances can be intractable. 
For example, consider a boolean ($\kset=\{0,1\}$) \pcsp
instance $I$ where all dyadic scores are~1,
and the monadic scores are 
$p_v\fcn 0 = 1$, $p_v\fcn 1 = z^{r_v}$,
with mutually incommensurable real values $r_v$.
Then each of the $2^n$ variable assignments corresponds to 
a different monomial in the partition function~$Z(I)$.
Since the output is of size $\Omega(2^n)$,
there is no hope of a time- or space-efficient algorithm.

However, some modest restrictions suffice to control the size of
the elements of the polynomial ring with which we are working
(including the output itself),
and thus the time- and space-efficiency of the various algorithms.

\subsection{Polynomially bounded instances}
The simplest well-behaved case is when an instance is over a single variable,
and all score polynomials are of low degree, with small coefficients.
This is easily formalized and generalized.

\newcommand{\FF}{\mathcal{F}}
\begin{definition}[Polynomially bounded]
A family $\FF$ of instances of \pcsp is \emph{polynomially bounded} if 
for an instance $I \in \FF$ having $n$ variables and $m$ clauses,
with respect to $m+n$:
\begin{itemize}
\item 
the number of formal variables is $O(1)$,
\item 
all powers are integers with absolute value $\Ostar1$, and
\item 
coefficients are integers of length $\Ostar1$,
or sums and products of coefficients are deemed to be elementary operations.
\end{itemize}
\end{definition}
For convenience we will speak of ``a polynomially bounded instance''.

\begin{lemma}
In any of the four algorithms described, over a fixed domain~$\kset$,
solving a polynomially bounded instance $I$ 
with $n$ variables and $m$ clauses,
each ring operation can be performed in 
time and space $\Ostar1$.
\end{lemma}

\begin{proof}
Say that there are $c$ variables, each with maximum degree~$\D$.
As a sum of products,
the partition function itself has maximum degree (in each variable)
at most $(m+n+1)\D$.
This is also true of any polynomial that arises in the course of
the computation. 
This can be verified directly by looking at the details of each algorithm.
Alternatively, note that the algorithms compute ``obliviously'' on
the polynomials (the operations performed depend on the constraint graph
but not on the polynomials),
the algorithms never subtract, 
and thus if on any input an algorithm ever produced terms of higher degree 
than the degree bound on the partition function, 
it would certainly do so when each input score polynomial is
$1+z+\cdots+z^\D$ (or the equivalent multivariate polynomial),
and (in the absence of any negative coefficients)
such terms would survive to appear in the partition function,
a contradiction.

Since each intermediate polynomial has maximum degree at most $(m+n+1)\D$,
it has at most $( (m+n+1)\D)^c = \Ostar1$ terms,
and two such polynomials can be multiplied with $\Ostar1$ 
integer operations.

If we wish to further break down the time as a function of 
the lengths of the integer coefficients,
each arises as a sum of at most $k^{m+n+1}$ products
of at most $m+n+1$ input coefficients,
and therefore has length at most $(m+n+1) \log k \cdot (m+n+1) = \Ostar1$
times that of the longest input coefficient.
Thus, adding and multiplying coefficients can also be done in time $\Ostar1$.
\end{proof}

\subsection{Pruned instances}
\label{prunes}
An alternative for a \pcsp in a single variable $z$
is to allow arbitrary (real) powers in the score polynomials,
but to avoid the blowup in our motivating example by 
demanding not the full partition function but just
its leading term (the one with the highest power of~$z$).
\begin{definition}[$z$-prunable]
An instance of \pcsp with $n$ variables and $m$ clauses 
(more precisely a family of instances) is
\emph{$z$-prunable} if,
with respect to $m+n$:
\begin{itemize}
\item the number of formal variables is $O(1)$,
\item $z$ may have arbitrary real powers,
\item the remaining variables $w_1,w_2,\ldots$ have
integer powers with absolute value no more than $\Ostar1$,
and
\item coefficients are nonnegative integers of length $\Ostar1$,
or coefficients are nonnegative reals 
and sums and products of coefficients are deemed to be elementary operations,
\end{itemize}
where the $O(\cdot)$ notation is with respect to $m+n$.
\end{definition}
Note that a \pcsp arising as the generating function of a \csp 
(see Section \ref{gfscsp})
is always $z$-prunable:
there is just a single formal variable~$z$, and all coefficients are~1.

\medskip

\noindent{\bf Pruned polynomial.}
Given a polynomial $p$ in one variable~$z$, we define the \emph{pruned
polynomial} $(p)_z$ to be the polynomial obtained by removing all but
the leading term.  If $p$ is a polynomial in variables $z, w_1,
w_2,\ldots$, we obtain $(p)_z$ by throwing away all terms $T$ such that
there is a term of the form $cTz^i$,
where $c>0$ is real, and $i>0$.  For instance,
\begin{align*}
(2z^2+3z+700+zw_1+z^2w_1+zw_2+z^{10}w_1w_2)_z
 &= 2z^2 +  z^2w_1+zw_2+z^{10}w_1w_2 .
\end{align*}

Pruning a partition function often preserves all the information of interest.
Consider for example a \pcsp $I$ arising from a simple maximization \csp, 
say weighted Max Cut. 
In the partition function $Z_I$, the leading term's power is 
the weight of a largest cut, and its coefficient the number of such cuts.
(For this informal discussion, we will simply multiply-count cuts 
that are symmetric to one another.)
With real edge weights, $Z_I$ can have exponentially many terms, 
but the weight of a largest cut, and the number of such cuts, 
is preserved in the pruned partition function $(Z_I)_z$,
which consists \emph{solely} of that leading term.

Somewhat more generally, consider a \csp where we wish to
maximize one (real-valued) parameter for one or more values of 
several other (integer-valued) parameters.
For example, the maximum bisection of an edge-weighted graph means 
maximizing the weight of cut edges, 
when the number of vertices in one partition is held to $n/2$.
Terms in the partition function $Z_I$ of the corresponding bivariate \pcsp 
have real powers of 
the edge-weighing variable~$z$, and
integral powers of the 
variable $w$ that counts vertices in partition~1.
Again (by definition) pruning to $(Z_I)_z$ preserves the maximum power 
of $z$ and the corresponding coefficient
for any fixed power of~$w$, 
so from $(Z_I)_z$ we may read off the weight of a maximum bisection 
and the number of such bisections, 
and indeed the weight and cardinality of maximum cuts of 
any specified sizes (say with exactly $n/3$ vertices in one partition).

\medskip

\noindent{\bf Pruning in the algorithms.}
The key point about pruning is that for polynomials $p$ and $q$ 
(in any set of variables) if coefficients are nonnegative then $(p+q)_z=(p_z+q_z)_z$ and $(pq)_z=((p)_z (q)_z)_z$.

With either of the reductive algorithms described,
the pruned partition function can be obtained efficiently,
first pruning the input instance, then pruning as we go.
(Correctness of this can be seen inductively,
working backwards from the last reduction.)
It is a trivial observation that for each reduction $R$
(of Type 0--3), and any instance~$I$,
$(R(I))_z=(R(I_z))_z$.
In particular, if we perform a full sequence of reductions 
and $z$-prune at every stage, 
we will end up with $(Z_I)_z$.

Pruning can also be done for tree decomposition-based
dynamic-programming algorithms.
Referring to the proof of Theorem~\ref{thm:treewidth},
we simply note that the pruned partition function can be obtained 
from pruned versions of $f_{B_1}$, $f'_{B_1}$, $f_{B_2}$, and $f'_{B_2}$.

This establishes the following lemma.

\begin{lemma}
Given a $z$-prunable PCSP instance $I$ 
with $n$ variables and $m$ clauses over a fixed domain~$\kset$,
the pruned partition function can be calculated by 
the reductive or dynamic programming algorithms
(see Theorems \ref{reductive}, \ref{random}, and~\ref{thm:treewidth})
with each ring operation taking time and space $\Ostar1$.
\end{lemma}

It is not likely that the \pcsp extension of Williams' algorithm 
can accommodate real-valued powers, even with pruning. 
One problem is the use of fast matrix multiplication as a ``black box'':
it is unlikely that we can prune within the multiplication algorithm.
Another piece of evidence against being able to accommodate 
real-valued \pcsp powers is that
in the \csp setting, Williams was unable to accommodate real-valued scores 
(the analogue of the \pcsp's powers),
and a solution for \pcsp would imply one for \csp.
Williams noted that one could work around real-valued scores 
by approximation methods,
for example multiplying all scores by a large constant
and then replacing each with the nearest integer,
and this can also be done in the \pcsp setting.

\section{Summary of PCSP algorithmic results}   \label{conclusions}
Combining the results in Sections~\ref{sec:Williams}--\ref{treewidth},
phrased in terms of the number of ring operations, 
with the bounds in Section~\ref{sec:ring}
on the complexity of these operations for PCSPs,
gives the following general conclusions.
Some particular results are presented in Section~\ref{keyresults}.

\begin{theorem} \label{ewilliams}
The \pcsp extension of Williams split-and-list algorithm
\cite{Williams}
solves any polynomially bounded \pcsp instance 
with $n$ variables over domain $\kset$
in time and space $\Ostar{k^{\omega n/3}}$,
where $\omega$ is the matrix-multiplication exponent
for the ring of polynomials over reals.
\end{theorem}
Parametrizing by vertices,
this exponential-space algorithm is the only 
known algorithm faster than $\Ostar{k^n}$
for \pcsp, just as it is for \csp.
In particular, it is the only algorithm efficient 
for dense \pcsp instances. 

\begin{theorem} \label{efaster}
The \pcsp extension of 
of Scott and Sorkin's reductive algorithm \cite{faster}
solves any polynomially bounded \pcsp instance
(or finds the pruned partition function of any prunable \pcsp instance)
with $n$ variables, $m$ clauses, and over domain $\kset$,
in time $\Ostar{k^{19m/100}}$ and space $\Ostar1$.
\end{theorem}
This algorithm is suitable for sparse instances
(if $m > (100/19) n$ then the naive $O(k^n)$ algorithm is better).
It is not as fast as dynamic programming (see Theorem~\ref{etree}), 
but unlike that and Williams' algorithm
it runs in polynomial space, and is potentially practical.

\begin{theorem} \label{elinear}
When $G = G(n,c/n)$ is an Erd\H os-R\'enyi random graph
with $c \leq 1+ \lambda n^{-1/3}$,
an extension of Scott and Sorkin's 
expected-linear-time \csp algorithm
\cite{linear}
solves any boolean \pcsp instance with constraint graph $G$
in expected time $\Ostar1 \exp(1+\lambda^3)$
and in space $\Ostar1$.
\end{theorem}
This polynomial-space, polynomial-expected-time algorithm
is of course the most efficient for sparse semi-random instances
(below the giant-component threshold or in its scaling window).

\begin{theorem} \label{etree}
Tree decomposition-based dynamic programming 
can solve any polynomially bounded \pcsp instance
(or find the pruned partition function of any prunable \pcsp instance)
with $n$ variables, $m$ clauses, over domain $\kset$,
and having treewidth $b-1$,
in time and space $\Ostar{k^b}$
or (since $b-1 \leq (13/75+o(1))m$, from \cite{faster,Kneis4})
time and space $\Ostar{k^{(13/75+o(1))m}}$.
\end{theorem}

\typeout{In CSP we had time $k^{13/75}$ but space $k^{1/9}$.
I don't remember why that was true there; could it be true here?}

This exponential-space algorithm is suitable for 
instances of small treewidth,
including sparse instances with $m<(75/13)n$.

\section{Constructing and sampling solutions} \label{sampling}
Wherever we can compute a \csp's maximum value, 
we can also produce a corresponding assignment; 
and wherever we can count assignments producing a given value, 
we can also do exact random sampling from these assignments. 
The method is standard, and we illustrate with sampling.
We construct our assignment one variable at a time, 
starting from the empty assignment.
Given a partial assignment
$\sigma_0:V_0\to\kset$, and a vertex $v\not\in I_0$, 
we calculate the partition functions
\begin{align}
Z_{I;\sigma_0\fcn i}
 =\sum_{\sigma \colon \sigma|_{V_0}=\sigma_0, \; \sigma(v)=i}I(\sigma) ,
 \label{Zsub}
\end{align}
and use these to determine the conditional distribution of 
$\sigma(v)$ given that $\sigma|_{I_0}=\sigma_0$.

This enables us to sample from a variety of distributions.  
For instance, we get the following result.

\begin{theorem} \label{esample}
Let $G$ be a graph with $m$ edges.  
Then in time $\Ostar{2^{19m/100}}$ and space $\Ostar1$
we can sample uniformly at random from the following distributions:
\begin{itemize}
\item maximum cuts, maximum bisections, minimum bisections
\item maximum independent sets, cliques of maximal size
\item independent sets of any fixed size, cliques of any fixed size.
\end{itemize}
In the same time, we can sample from the equilibrium (Gibbs) distribution 
of the Ising model with any fixed interaction strength 
and external magnetic field.
\end{theorem}

A similar result holds for edge- and/or vertex-weighted graphs, 
except that we sample at random only from optimal assignments.
Many other problems can be expressed in this framework. 
For example, we can
count and sample proper $k$-colorings 
in time $\Ostar{k^{19m/100}}$ and space $\Ostar 1$,
or in time and space $\Ostar{k^{(13/75+o(1))m}}$.

If we wish to exhibit just 
one optimal assignment deterministically, 
we can do so by a small modification of the sampling approach above: 
at each step, assign the next vertex $v$ the smallest $i$ such that 
$\sigma_0\fcn i$ can be extended to an optimal assignment 
(this can be read off from $Z_{I;\sigma\fcn i}$
given by~\eqref{Zsub}). 

\section{Applications and Conclusions} \label{keyresults}
The PCSP / RCSP framework encompasses many problems that are not in CSP, 
yet can be solved by extensions of any general CSP algorithm we know of, 
with essentially equal efficiency.
For several problems, this yields the best algorithms known.

\subsection{Graph bisection}
Jansen, Karpinski, Lingas and Seidel \cite{Karpinski} showed
how to use dynamic programming based on tree decomposition to
find a maximum or minimum bisection of a graph. 
Combined with the bounds from \cite{faster,Kneis4}
on the treewidth of a graph with $m$ edges, 
this gave an algorithm with a time and space bound of 
$\Ostar{2^{(13/75+o(1))m}}$.

Until now, no polynomial-space algorithm was known for this problem
other than the naive enumeration running in time $\Ostar{2^n}$.
Once bisection is phrased as a PCSP,
our results (Theorem~\ref{efaster}) immediately imply that
in time $\Ostar{2^{19m/100}}$ and polynomial space,
we can solve maximum and minimum bisection,
count maximum and minimum bisections, and 
indeed count bisections of all sizes.
We can do this also in weighted bisection models,
as long as they are polynomially bounded or $z$-prunable.

Our results (Theorem~\ref{ewilliams}) also mean that dense instances
of bisection can be solved by Williams' algorithm,
in time and space $\Ostar{2^{\omega n/3}}$, 
again yielding the full generating function of cuts.

\subsection{Cliques and Independent Sets}

Maximum Independent Set is in Max 2-CSP and has been studied extensively,
but, surpisingly, PCSP can offer an advantage here.
For simple MIS, for any graph with average vertex degree $d$, 
all four of our general-purpose PCSP algorithms are dominated 
either by \cite{BourgeoisEP2008} 
(for average degree $\leq 3$, time $\Ostar{2^{0.1331 n}}$)
or by \cite{FGK} (time $\Ostar{2^{0.288 n}}$, otherwise).
However, for weighted MIS, counting MIS, or weighted counting MIS,
the fastest algorithm we are aware of is the vertex-parametrized 
$\Ostar{2^{0.3290 n}}$ algorithm of F\"urer and Kasiviswanathan \cite{FK},
based on an improved analysis of an 
algorithm of Dahll\"of, Jonsson, and Wahlstr\"om \cite{dahllofTCS}.
For any of these problems
the $\Ostar{2^{19m/100}}$  reductive algorithm of Theorem~\ref{efaster} 
is the fastest polynomial-space algorithm known 
if the average degree is below about~$3.46$,
and the $\Ostar{2^{(13/75+o(1))m}}$
exponential-space dynamic-programming algorithm 
of Theorem~\ref{etree} the fastest algorithm known 
if the average degree is below about~$3.79$.

We noted earlier that Maximum Clique is not in the class 2-CSP, 
because constraints are given by non-edges, not edges, 
but that it is in the class PCSP.
This is nice, but does not currently offer any algorithmic advantage.
If $G$ has average degree $d=o(n)$, the following simple algorithm 
has running time $\Ostar{2^{o(n)}}$, dominating all other 
algorithms under consideration here.
For any $k$ of our choosing,
any clique either is contained in the $\leq n/k$-order subgraph induced 
by vertices of degree $\geq kd$, 
which is checkable in time $\Ostar{2^{n/k}}$,
or contains a vertex of degree $\leq kd$, 
which is checkable in time~$\Ostar{2^{kd}}$.
Taking $k=\sqrt{n/d}$
gives running time $\Ostar{2^{\sqrt{nd}}}$.
If $d$ is not $o(n)$ then certainly $d=\omega(1)$.
This rules out edge-parametrized algorithms, 
leaving us only Williams' algorithm,
which is less efficient than applying a vertex-parametrized 
MIS algorithm such as that of \cite{FK} to the complement graph~$\Gbar$.

\subsection{Ising partition function}
The Ising partition function,
a canonical object in statistical physics, is a PCSP, 
and can be solved by any of the algorithms presented here.
We know of no other general algorithm to find the partition function
other than naive enumeration over all $2^n$ configurations.

\subsection{Potts and other $q$-state models}
The Potts ($q$-coloring) partition function can trivially be evaluated
in time $\Ostar{q^n}$,
but it is by no means obvious that one can remove the dependence on~$q$.
A breakthrough result of 
Bj\"orklund, Husfeldt, Kaski, and Koivisto \cite{FOCS2008} 
(see also
\cite{FOCS2006B,FOCS2006K,BjorklundSIAM2008,STOC2007,Bjorklund2006,Husfeldt2008})
shows how to evaluate the Potts partition function 
(and therefore also the Tutte polynomial)
using the {Inclusion--Exclusion method},
in time $\Ostar{2^n}$ in exponential space,
or time $\Ostar{3^n}$ in polynomial space.

Our PCSP model includes the Potts model 
and more general models where the ``energy'' is not simply a function 
of equality or inequality of neighboring colors. 
Our algorithms unfortunately do have have exponential dependence on~$q$, 
but with the greater generality of PCSP this may be unavoidable:
as noted in \cite{FOCS2008},
Traxler \cite{TraxlerIWPEC2008} shows that if the $q$-state Potts model 
is generalized even to 2-CSPs on variables with $q$ states, 
exponential dependence on $q$ cannot be avoided,
at least in the vertex-parametrized case,
assuming the Exponential Time Hypothesis.

\subsection{Conclusions} 
As we have shown, the PCSP formulation is clean 
and offers quantifiable advantages for a wide range of problems. 
However, the greatest benefit is its broad applicability. 
For example, exponential time algorithms have not previously been considered 
for problems such as judicious partitioning, 
and for virtually any form of this problem such algorithms 
now follow instantly from its membership in PCSP.  
Moreover, when solving optimization problems coded into PCSP form, 
we automatically get solutions to the corresponding counting problems.

Finally, we note that algebraic formulations appear to be playing 
a newly important role in the development of 
exponential-time algorithms for CSPs and related problems. 
Our work along these lines began with \cite{CountingArxiv-v1}, 
and we have already discussed Koivisto's sum-of-products variation 
on Williams' split-and-list algorithm \cite{Koivisto}. 
The methods leading to the $\Ostar{2^n}$ Tutte polynomial computation 
\cite{FOCS2008} work in a fairly general algebraic setting, 
and a new algorithm of Williams \cite{WilliamsPath} 
that finds a $k$-path in an order-$n$ graph in time $\Ostar{2^k}$
works in an algebra carefully chosen for properties including making
walks other than simple paths cancel themselves out.
PCSPs should be considered one piece of this new algebraic tool kit.

\providecommand{\bysame}{\leavevmode\hbox to3em{\hrulefill}\thinspace}
\providecommand{\MR}{\relax\ifhmode\unskip\space\fi MR }
\providecommand{\MRhref}[2]{%
  \href{http://www.ams.org/mathscinet-getitem?mr=#1}{#2}
}
\providecommand{\href}[2]{#2}


\begin{thebibliography}{KMRR09}

\bibitem[AHE94]{DAM}
S.~Arnborg, S.T. Hedetniemi, and A.~Proskurowski (Eds.), Discrete Applied
  Mathematics \textbf{54} (1994), no.~2-3, special issue.

\bibitem[AP89]{Arnborg}
S.~Arnborg and A.~Proskurowski, \emph{Linear time algorithms for {NP}-hard
  problems restricted to partial $k$-trees}, Discrete Applied Mathematics
  \textbf{23} (1989), no.~1, 11--24.

\bibitem[BEP08]{BourgeoisEP2008}
Nicolas Bourgeois, Bruno Escoffier, and Vangelis~Th. Paschos, \emph{An
  {$O^*(1.0977^n)$} exact algorithm for max independent set in sparse graphs},
  IWPEC (Martin Grohe and Rolf Niedermeier, eds.), Lecture Notes in Computer
  Science, vol. 5018, Springer, 2008, pp.~55--65.

\bibitem[BG05]{Bulatov}
Andrei Bulatov and Martin Grohe, \emph{The complexity of partition functions},
  Theoret. Comput. Sci. \textbf{348} (2005), no.~2-3, 148--186.

\bibitem[BH06a]{Bjorklund2006}
Andreas Bj{\"o}rklund and Thore Husfeldt, \emph{Exact algorithms for exact
  satisfiability and number of perfect matchings}, Proceedings of the 33nd
  International Colloquium on Automata, Languages and Programming (ICALP 2006),
  Lecture Notes in Computer Science, vol. 4051, Springer, 2006, pp.~548--559.

\bibitem[BH06b]{FOCS2006B}
\bysame, \emph{Inclusion-exclusion algorithms for counting set partitions},
  Proceedings of the 47th {IEEE} Symposium on Foundations of Computer Science
  (FOCS 2006), 2006, pp.~575--582.

\bibitem[BH08]{Husfeldt2008}
Andreas Bj{\"o}rklund and Thore Husfeldt, \emph{Exact algorithms for exact
  satisfiability and number of perfect matchings}, Algorithmica \textbf{52}
  (2008), no.~2, 226--249.

\bibitem[BHK]{BjorklundSIAM2008}
Andreas Bj{\"o}rklund, Thore Husfeldt, and Mikko Koivisto, \emph{Set
  partitioning via inclusion-exclusion}, {SIAM} Journal on Computing, special
  issue dedicated to selected papers from FOCS 2006, to appear.

\bibitem[BHKK07]{STOC2007}
Andreas Bj{\"o}rklund, Thore Husfeldt, Petteri Kaski, and Mikko Koivisto,
  \emph{Fourier meets {M}\"obius: fast subset convolution}, {P}roceedings of
  the 39th {A}nnual {ACM} {S}ymposium on {T}heory of {C}omputing (STOC 2007),
  ACM, New York, 2007, pp.~67--74.

\bibitem[BHKK08]{FOCS2008}
Andreas Bj{\"o}rklund, Thore Husfeldt, Petteri Kaski, and Mikko Koivisto,
  \emph{Computing the {T}utte polynomial in vertex-exponential time},
  Proceedings of the 49th {IEEE} Symposium on Foundations of Computer Science
  (FOCS 2008), 2008.

\bibitem[Bod96]{Bodlaender}
Hans~L. Bodlaender, \emph{A linear-time algorithm for finding
  tree-decompositions of small treewidth}, SIAM J. Comput. \textbf{25} (1996),
  no.~6, 1305--1317.

\bibitem[BS99]{BScomb}
B{\'e}la Bollob{\'a}s and Alexander~D. Scott, \emph{Exact bounds for judicious
  partitions of graphs}, Combinatorica \textbf{19} (1999), no.~4, 473--486.

\bibitem[CW90]{Coppersmith-Winograd}
Don Coppersmith and Shmuel Winograd, \emph{Matrix multiplication via arithmetic
  progressions}, J. Symbolic Comput. \textbf{9} (1990), no.~3, 251--280.

\bibitem[DGP06]{Dyer}
Martin~E. Dyer, Leslie~Ann Goldberg, and Mike Paterson, \emph{On counting
  homomorphisms to directed acyclic graphs.}, ICALP (1), 2006, pp.~38--49.

\bibitem[DJW05]{dahllofTCS}
Vilhelm Dahll{\"o}f, Peter Jonsson, and Magnus Wahlstr{\"o}m, \emph{Counting
  models for 2{SAT} and 3{SAT} formulae}, Theoret. Comput. Sci. \textbf{332}
  (2005), no.~1-3, 265--291.

\bibitem[DP87]{Dechter87}
R.~Dechter and J.~Pearl, \emph{Network-based heuristics for
  constraint-satisfaction problems}, Artif. Intell. \textbf{34} (1987), no.~1,
  1--38.

\bibitem[DP89]{Dechter89}
Rina Dechter and Judea Pearl, \emph{Tree clustering for constraint networks
  (research note)}, Artif. Intell. \textbf{38} (1989), no.~3, 353--366.

\bibitem[FGK06]{FGK}
Fedor~V. Fomin, Fabrizio Grandoni, and Dieter Kratsch, \emph{Measure and
  conquer: a simple {$O(2^{0.288 n})$} independent set algorithm}, Proceedings
  of the 17th annual {ACM-SIAM} symposium on Discrete Algorithms (SODA 2006)
  (New York, NY, USA), ACM, 2006, pp.~18--25.

\bibitem[FH06]{fomin}
Fedor~V. Fomin and Kjartan H{\o}ie, \emph{Pathwidth of cubic graphs and exact
  algorithms}, Inform. Process. Lett. \textbf{97} (2006), no.~5, 191--196.

\bibitem[FK05]{FK}
Martin F\"urer and Shiva~Prasad Kasiviswanathan, \emph{Algorithms for counting
  {2-SAT} solutions and colorings with applications}, Tech. Report~33,
  Electronic Colloquium on Computational Complexity, 2005.

\bibitem[HN04]{Hell}
Pavol Hell and Jaroslav Ne{\v{s}}et{\v{r}}il, \emph{Graphs and homomorphisms},
  Oxford Lecture Series in Mathematics and its Applications, vol.~28, Oxford
  University Press, Oxford, 2004.

\bibitem[JKLS05]{Karpinski}
Klaus Jansen, Marek Karpinski, Andrzej Lingas, and Eike Seidel,
  \emph{Polynomial time approximation schemes for max-bisection on planar and
  geometric graphs}, SIAM J. Comput. \textbf{35} (2005), no.~1, 110--119
  (electronic).

\bibitem[KMRR09]{Kneis4}
Joachim Kneis, Daniel M{\"o}lle, Stefan Richter, and Peter Rossmanith, \emph{A
  bound on the pathwidth of sparse graphs with applications to exact
  algorithms}, {SIAM} J. Discrete Math., vol.~23, SIAM, 2009, pp.~407--427.

\bibitem[Koi06a]{FOCS2006K}
Mikko Koivisto, \emph{An {$O^*(2^n)$} algorithm for graph colouring and other
  partitioning problems via inclusion-exclusion}, Proceedings of the 47th
  {IEEE} Symposium on Foundations of Computer Science (FOCS 2006), 2006,
  pp.~583--590.

\bibitem[Koi06b]{Koivisto}
Mikko Koivisto, \emph{Optimal 2-constraint satisfaction via sum-product
  algorithms}, Inform. Process. Lett. \textbf{98} (2006), no.~1, 24--28.

\bibitem[MP01]{Monien}
Burkhard Monien and Robert Preis, \emph{Upper bounds on the bisection width of
  3- and 4-regular graphs}, Mathematical foundations of computer science, 2001
  (Mari\'ansk\'e L\'azn\u e), Lecture Notes in Comput. Sci., vol. 2136,
  Springer, Berlin, 2001, pp.~524--536.

\bibitem[Ree92]{Reed}
Bruce~A. Reed, \emph{Finding approximate separators and computing tree width
  quickly}, STOC, ACM, 1992, pp.~221--228.

\bibitem[RS95]{robertsonpath}
N.~Robertson and P.~D. Seymour, \emph{{Graph minors {XIII}: The disjoint path
  problem}}, J. Combinatorial Theory (Series B) \textbf{63} (1995), 65--110.

\bibitem[Sco05]{Sbcc}
Alexander~D. Scott, \emph{Judicious partitions and related problems}, Surveys
  in combinatorics 2005, London Math. Soc. Lecture Note Ser., vol. 327,
  Cambridge Univ. Press, Cambridge, 2005, pp.~95--117.

\bibitem[SS06a]{CountingArxiv-v1}
Alexander~D. Scott and Gregory~B. Sorkin, \emph{Generalized constraint
  satisfaction problems}, Tech. Report cs:DM/0604079v1, arxiv.org, April 2006,
  See {http://\allowbreak arxiv.org/\allowbreak abs/\allowbreak
  cs.DM/\allowbreak 0604079}.

\bibitem[SS06b]{linear}
\bysame, \emph{Solving sparse random instances of {Max Cut} and {Max 2-CSP} in
  linear expected time}, Comb. Probab. Comput. \textbf{15} (2006), no.~1-2,
  281--315.

\bibitem[SS07]{faster}
Alexander~D. Scott and Gregory~B. Sorkin, \emph{Linear-programming design and
  analysis of fast algorithms for {M}ax 2-{CSP}}, Discrete Optim. \textbf{4}
  (2007), no.~3-4, 260--287.

\bibitem[Tra08]{TraxlerIWPEC2008}
Patrick Traxler, \emph{The time complexity of constraint satisfaction},
  Proceedings of the 3rd International Workshop on Parameterized and Exact
  Computation (IWPEC 2008), Lecture Notes in Computer Science, vol. 5018,
  Springer, 2008, pp.~190--201.

\bibitem[Wil04]{Williams}
Ryan Williams, \emph{A new algorithm for optimal constraint satisfaction and
  its implications}, Proc.\ 31st International Colloquium on Automata,
  Languages and Programming (ICALP), 2004.

\bibitem[Wil08]{WilliamsPath}
\bysame, \emph{Finding paths of length $k$ in {$O^*(2^k)$} time}, Tech. Report
  cs.DS/0807.3026v3, arxiv.org, November 2008, See {http://\allowbreak
  arxiv.org/\allowbreak abs/\allowbreak cs.DM/\allowbreak 0604080}.

\end{thebibliography}
\end{document}